# Two Approaches for the Passive Charge Management of Contactless Test Masses

S. Wang[1,2], S. Saraf[1], J. Lipa[3], D. Yadav[1], S. Buchman[3]
[1]SN&N Electronics, Inc., 1846 Stone Ave, San Jose, CA 95125
[2]Hainan Tropical Ocean University, Sanya, China
[3]Stanford University, W.W. Hansen Laboratory, Stanford CA 94305

Free floating Test Masses (TM) of inertial reference instruments accumulate charge mainly through the triboelectric effect during separation from their housings and, if in the space environment, from cosmic radiation. These charges will degrade the accuracy of high sensitivity accelerometers and drag-free sensors. We demonstrate in ground testing two passive bipolar charge management systems using photoelectrons emitted from gold coated surfaces under illumination by Ultraviolet Light Emitting Diodes (UV-LEDs) with 255 nm, 275 nm, and 295 nm central wavelengths. The first method uses fast photoelectrons, generated by two 255 nm UV-LEDs with adjustable-intensity (through fine-tuning of their excitation currents) and illuminating the TM and its housing respectively. A second technique uses slow photoelectrons generated by one UV-LED, of either 275 nm or 295 nm, directed at the TM. Fast and slow electrons are defined as having kinetic energies after photoemission above and below $eV_{TM}^{max}$, where $V_{TM}^{max}$ is the maximum allowable potential required for normal operation of the TM. In its optimized configurations and following an exposure of < 30 sec to UV, the fast-photoelectron control system converges to zero TM potential from ± 1 V with a drift of ≈ 1.5 mV/day. The slow-photoelectron system requires about 5 minutes to converge to zero TM potential from 100 mV, with a similar drift of ≈ 2.0 mV/day. For reference, $V_{TM}^{max} \cong 80mV$ for the LISA and LISA pathfinder sensors. Insight into the slow photoelectrons charge control method is provided by our simple gaussian model of the work function for gold films (that approximates their quasi-triangular measured shape): 4.6 eV center, 0.2 eV standard deviation, and 4.30 eV and 4.90 eV cutoffs. The longest wavelength (lowest energy) photoemission is obtained with the 295 nm LED (11 nm FWHM) at about 285 nm (4.35 eV), where only about 3.5% of the photons have an energy above the photoelectric threshold. While the two-source UV fast photoelectron method is a modification of the LISA Pathfinder approach, the slow photoelectron method is a new passive charge management technique. These two passive techniques were developed for instruments without sensing and activation systems. For instruments with electric fields surrounding the TM and more complex geometries additional adaptations are required.

## I. Introduction

Inertial sensors using bulk TMs achieve their best performance in the space environment. Examples of missions are the DISturbance COmpensation System (DISCOS)[1], the Gravity Recovery And Climate Experiment (GRACE)[2], the Gravity field and steady-state Ocean Circulation Explorer (GOCE)[3], the Relativity mission Gravity Probe B (GP-B)[4], the Pathfinder LPF[5] for the Laser Interferometer Space Antenna (LISA)[6] mission, and the Micro-Satellite à traînée Compensée pour l'Observation du Principe d'Equivalence (MICROSCOPE)[7]. Electric charges on the TM interact with internal and external electromagnetic fields, degrading the performance of the instrument and are therefore limited to ≤10 pC for GP-B[8] and to ≤ 3 pC for LPF and LISA[9,10,11]. DISCOS discharged its TM by bringing it in contact with its housing while the GRACE, GOCE, and MICROSCOPE TMs have fine gold wires grounding them to their housings.

UV generated photoelectrons were used for GP-B[12,13] and LPF[14] (and planned for LISA[15], Taiji[16], and TianQin[17]) and successfully demonstrated this charge management technique. The photoelectrons were generated by the 254 nm UV line of Hg discharge lamps, while force modulation was the method of TM charge measurement[12]. A different system, using ionized gas[18], has been implemented for the suspended mirrors of the ground-based Laser Interferometer Gravitational Observatory (LIGO)[19].

Development of LEDs at ≤ 255 nm[20] has motivated the replacement of Hg lamps with LEDs as UV sources for these applications. UV sources powered by LEDs are lighter, have longer lifetime, lower power consumption, and superior radiation resistance[21,22] compared to Hg lamps. Using the GP-B hardware for comparison, we estimate that the replacement of Hg lamps with LEDs will reduce the overall power, weight, and volume requirements for the charge management system by more than 50%. Furthermore, LEDs can be modulated at frequencies well above the operational bandwidth of the inertial sensors, thus reducing disturbances and allowing for AC charge management[23,24]. A collaboration of Stanford University, NASA and KACST[25] has successfully flown a technology demonstration of charge management with UV-LEDs[26] - the UV-LED Mission launched in June 2014 on the SaudiSat-4 small satellite[27].

In this paper we present data obtained from laboratory measurements that validate two techniques of passive charge management, i.e., requiring none or infrequent (≥1 week) measurement of the TM charge. These techniques significantly simplify the charge management process, improve the overall instrument performance, and reduce its mass and complexity. All inertial sensors, and in particular those not requiring electrostatic forcing (spherical TM for example), will greatly profit from a charge management system that does not require charge measurement and can be activated infrequently over short periods. While demonstrated in this work with DC excitation currents for the LEDs, the two methods work equally well in the AC mode[23], when no disturbing electric fields are present, thus bringing their frequencies well above the operational band of the instruments under consideration. In the LISA configuration, the injection and actuation voltages generate electric fields that are higher than the photoelectron energies, thus requiring synchronization with the activation of the UV-LEDs or similar methods of operation. We refer to the techniques as the Fast PhotoElectron (FPE) and Slow PhotoElectron (SPE) methods and describe them in detail below. The FPE method independently validates the work described by F. Yang et al.[28] for a generic configuration and expands on it by explicitly showing the dependence of the system parameters on vacuum preparation and temperature. The SPE approach is by design insensitive to system parameters for an equilibrium TM potential $|V(TM)| < 10$mV.

## II. Experimental Setup

Figure 1 shows the experimental setup (schematic and photograph) consisting TMs in the form of two plates made of gold-coated sheets of aluminum, of 14 cm by 9.6 cm, with 2.5 cm-separation, and labeled $TM^a$ and $TM^b$. The plates are approximately symmetrically located within the housing, allowing interchangeability in their function. Each plate is freely suspended by a 0.8 mm diameter gold wire that connects it, through a hole in the EMI cage and an insulated feedthrough through the vacuum enclosure (kept at $10^{-3}$ to $10^{-4}$ Pa), to the measurement instrumentation. The measurement instrumentation for each plate can be switched between the following four modes: $V_G^{a/b}$ a connection to ground, $V_B^{a/b}$ a voltage source that supplies a DC bias potential, a Keithley model 6485 pico-ammeter that measures the TM current to ground $I(TM^{a/b})$, and a Keithley model 6517 electrometer that measures the TM potential to ground $V(TM^{a/b})$. The pico-ammeter and/or the electrometer are used on one plate at a time.

**Table 1. Mutual capacitances for system**

| Cap. (pF) | $TM^a$ | $TM^b$ | EMI encl. | EMI encl + cabling |
|---|---|---|---|---|
| $TM^a$ | N/A | 9.3 | 18.9 | 169 |
| $TM^b$ | 9.3 | N/A | 16.6 | 251 |

LEDs, labeled $LED^A$ and $LED^B$, are mounted on the EMI enclosure. In order for the two parallel plates to simulate the TM and its housing, the photons should be reflecting between them, thus requiring the use of light-pipes with their ends mounted against the LEDs to guide the UV photons to illuminate only the opposing TMs through the center holes of the plates adjacent to the LEDs. The light-pipes[29] are of fused silica with hexagonal cross section, 2 mm inradius, 10 cm length, and a numerical aperture of $\cong 0.5$ (half angle $\cong 30°$), resulting in a nominal 29 mm diameter illuminated spot on the opposite TM. For this setup less than 2% of the UV power will be lost in the opposite light-pipe. This configuration is a nominal representation of the parallel plate configuration of many inertial sensors, including LISA and GRACE, though beam incidence angles might be different. The EMI enclosure functions as the ground of the system and is housed inside a vacuum chamber. Table 1 gives the mutual capacitances in pF of the components of the system, used for calculating the charges accumulated during the experiments; the last column shows the capacitance of the TMs to the EMI enclosure plus the coaxial cables to the instrumentation.

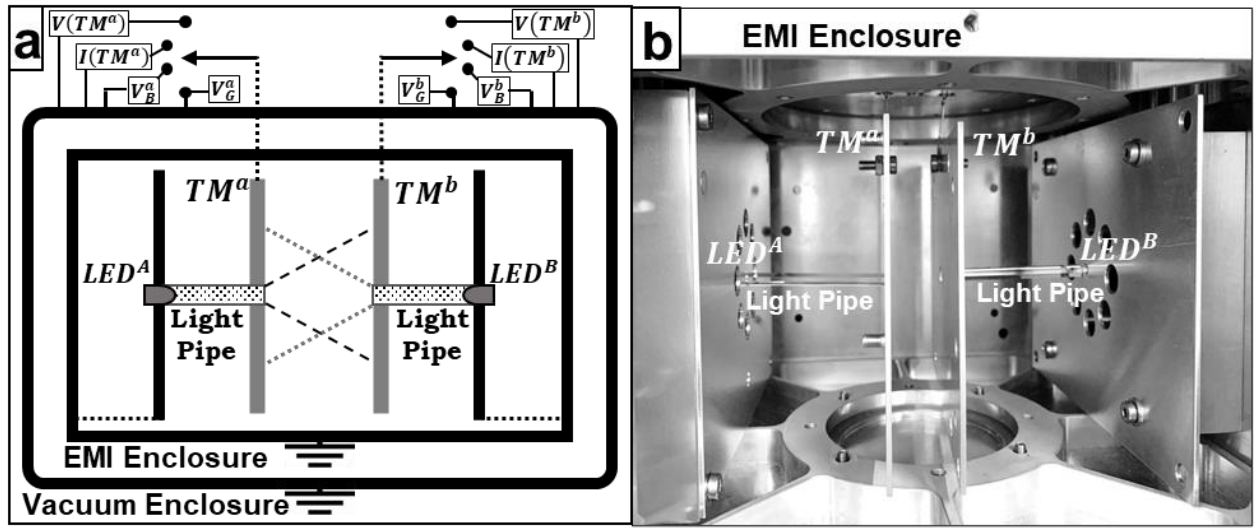

**Figure 1. Two-plate experimental setup: a) schematic, b) photograph.**

For characterization of their reflectivity and photoemission efficiency, gold film samples (similar to the thin films on the TM and the EMI housing) were deposited on 1 cm by 2 cm slides from a 99.9% purity gold target in a sputtering system. A 1000 Å gold film thickness was chosen by requiring that the ratio of photoelectrons emitted at the coating interface to the substrate to those emitted from the gold film be negligeable[30]. For reference, the electron mean free path in the gold film is about 70 Å[31]. Balakrishnan[30] measured the reflection coefficient $R$ and quantum efficiency $\Psi^*$ (defined here as photoelectrons per incident UV photon) for these gold film samples, that are similar to our TM coatings, to be $R = 0.17$ and $\Psi^* = 3.4 \times 10^{-7} \pm 10\% \frac{\text{photoelectrons}}{\text{incident photons}}$.

The TMs, the characterization samples, and the surfaces facing them were cleaned with 99% pure isopropyl alcohol just before pump-down, in order to improve the reproducibility of the results. The vacuum chamber was evacuated to between 10⁻³ Pa to 10⁻⁴ Pa before data collection, with the vacuum improving over week-long periods. The vacuum gauge was found to affect the measurement results significantly, most likely due to ions generated in the residual gas and was turned off before the start of the measurements. Data sampling interval was typically one second. Table 2 shows the characteristics of the UV-LEDs used in this work. LED power spectra resemble a slightly asymmetric triangle; see for example the measured 255 nm LED spectra in references 20, 30, 32. In Fig. 2 we approximate these power spectra as normalized to unity gaussian distributions, centered at $E_c$, with cutoffs at $\pm 2.5 \times \sigma$. The horizontal axis is converted from wavelength to energy. $E^{max/min} \equiv E_c \pm 2.5 \times \sigma(E)$ was selected at ≅4% of maximum amplitude to approximate the aforementioned triangle spectra. The shaded last row of the table gives the parameters of a simplified model that approximates the energy spectrum for the TM work function and will be addressed in the next section.

**Table 2. Characteristics of LEDs used in the experiment**

| LED | Model | $\lambda_C$ (nm) | FWHM (nm) | $E_c$ (eV) | $\sigma(E)$ (meV) | $P$ (µW/mA) | $E^{max}_{LED\_\lambda}$ (eV) | $E^{min}_{LED\_\lambda}$ (eV) |
|---|---|---|---|---|---|---|---|---|
| LED_255 | TUD59H1B[33] | 255±1 | 11±1 | 4.86±0.02 | 81±8 | 16±1 | 5.06 | 4.66 |
| LED_275 | OPTAN-275J-BL[34] | 275±5 | 11±1 | 4.51±0.10 | 70±7 | 15±5 | 4.68 | 4.34 |
| LED_295 | MTSM295UV-F1120S[35] | 295±5 | 11±1 | 4.20±0.07 | 60±6 | 100±10 | 4.35 | 4.05 |
| LED_310 | MTE-F13[35] | 310±5 | 10±1 | 4.00±0.06 | 55±5 | 40±4 | 4.14 | 3.86 |
| **TM** | **N/A** | **270** | **23.5.** | **4.60** | **170** | **N/A** | **4.90** | **4.30** |

### III. Photoemission Considerations

LPF and LISA have stringent requirements for the allowable TM charge[9,10], $Q_{TM}^{max} \leq 3\text{pC}$, that for $C(LISA) \approx 38\text{pF}$, gives the maximum allowed TM potential $V_{TM}^{max} \cong 80$ mV. The two approaches to passive charge management we describe here use photoelectrons in the 'fast' and 'slow' energy regimes respectively. We require that $\mathbb{K}^{FPE}$, the kinetic energy used in the FPE method, should be $\mathbb{K}^{FPE}/e > V_{TM}^{max}$, where $e$ is the elementary charge. For the SPE method $\mathbb{K}^{SPE}/e < V_{TM}^{max}$ and the maximum kinetic energy $\mathbb{K}^{max}(\lambda, TM)$ of photoelectrons, for an LED of wavelength $\lambda$ and a specific TM is:

$$\mathbb{K}^{max}(\lambda, TM) \leq E_\lambda^{max} - E_{TM}^{min} \quad (1)$$

$E_\lambda^{max}$ is the maximum of the energy distribution of the UV photons $(+ 2.5\sigma)^{20}$ and $E_{TM}^{min}$ is the minimum of the work function distribution of the TM chosen at -1.75$\sigma$. The work function for the gold coating of the TM has a quasi-triangular energy distribution (see for example reference 36) that we model as a gaussian with cutoffs. We choose $\sigma = 170$meV and the cutoffs at $\pm 1.75\sigma$ to best represent the shape and width of the work function. Photons emitted by $LED^A$ or $LED^B$ reflect back and forth between $TM^a$ and $TM^b$ with a reflection coefficient[30] $R \cong 0.17\,R$. This section addresses only conditions with no applied biases to the TMs.

In the FPE regime the photoelectrons are ballistic. After their initial illuminations and subsequent reflections from each TM, $LED^A$ and $LED^B$ generate the photoelectric currents $I^{Aa}$ and $I^{Ba}$ at $TM^a$ that, with no applied biases, are given by:

$$I^{Aa} = \Phi^A \sum_{i=1}^{n\to\infty} \left[ \left(\Psi_i^{Aa} - \Psi_i^{Ab} R_i^{Aa}\right) \prod_{j=0}^{i-1} R_j^{Ab} R_j^{Aa} \right] \quad (2)$$

$$I^{Ba} = \Phi^B \sum_{i=1}^{n\to\infty} \left[ \left(\Psi_i^{Ba} R_i^{Bb} - \Psi_i^{Bb}\right) \prod_{j=0}^{i-1} R_j^{Ba} R_j^{Bb} \right] \quad (3)$$

where $\Phi^A$ and $\Phi^B$ are the total photon fluxes (photons/s) of $LED^A$ and $LED^B$, and $\Psi_i^{Cc}$ and $R_i^{Cc}$ are the quantum efficiencies (current per incident photon flux) and reflectivities at the $i^{\text{th}}$ illumination from $LED^C$ on $TM^c$. Here we use $C = A$ or $B$ to refer to the LED, and $c = a$ or $b$ to refer to the TM. Here we assume for generality that unequal LED illumination geometry, specular reflection, and non-uniformity of the TM gold coatings cause the constants $\Psi_i^{Cc}$ and $R_i^{Cc}$ to depend on $C$ and $c$ and vary with $i$. However, the energy distribution of the UV photons is not considered explicitly. Because $\mathbb{K}^{FPE}/e > V_{TM}^{max}$, we can ignore the effect of $V_{TM}^{max}$ on $\Psi_i^{Cc}$. We define the summations in equations 2 and 3 as:

$$\mathbb{C}^{A(a,b)} \equiv \sum_{i=1}^{n\to\infty} \left[ \left(\Psi_i^{Aa} - \Psi_i^{Ab} R_i^{Aa}\right) \prod_{j=0}^{i-1} R_j^{Ab} R_j^{Aa} \right] \quad (4)$$

$$\mathbb{C}^{B(a,b)} \equiv \sum_{i=1}^{n\to\infty} \left[ \left(\Psi_i^{Ba} R_i^{Bb} - \Psi_i^{Bb}\right) \prod_{j=0}^{i-1} R_j^{Ba} R_j^{Bb} \right] \quad (5)$$

where $\mathbb{C}^{A(a,b)}$ and $\mathbb{C}^{B(a,b)}$, albeit complex to calculate, are constants. The total current generated by $LED^A$ and $LED^B$ at $TM^a$ is then given by:

$$I^{(A+B)a} = I^{Aa} + I^{Ba} = \Phi^A \mathbb{C}^{A(a,b)} + \Phi^B \mathbb{C}^{B(a,b)} \quad (6)$$

$I^{(A+B)a}$ is linear in $\Phi^A$ and $\Phi^B$ and, for the FPE technique, can be nulled, allowing us to set $V(TM) = 0V$ by adjusting $\Phi^A/\Phi^B$. This can be achieved by a variety of methods including the modulation of the LEDs' duty cycles or as demonstrated in this work, by varying the ratio of the excitation currents of the $LED^A$ and $LED^B$. The photon flux of the UV-LEDs is linearly dependent on their excitation current[33], thus facilitating the fine-tuning of $\Phi^A/\Phi^B$.

For better physical insight into the process and to facilitate actual estimates of the photoelectric current flowing in the system, we apply the simplifying approximations of $TM^a$ and $TM^b$ having a) equal and constant quantum efficiencies and reflectivity coefficients over their entire surfaces, and b) balanced illumination geometry. Equation 6 then reduces to:

$$I^{(A+B)a} = [\Phi^A(\Psi^a - \Psi^b R^a) + \Phi^B(\Psi^a R^b - \Psi^b)]/(1 - R^a R^b) \quad (7)$$

Equation 7 indicates that in certain configurations it might not be possible to null the TM current. Proper geometrical design and surface preparation are required to ensure that both right-side terms do not have the same sign.

The question of surface work functions and quantum efficiencies as functions of photon energy, and of photoelectron emission spectra from practical gold surfaces is complex and not fully understood[36]. For the SPE method we consider a simplified model supporting the approach of using photoelectrons with $\mathbb{K}^{SPE}/e < V_{TM}^{max}$. We assume a gaussian distribution for the energies of the photoelectric emission sites of the TM centered at 4.60 eV. $E_{TM}^{max/min}(TM) \equiv E_c \pm 1.75 \times \sigma(E)$ are at 4.30 eV and 4.90 eV[36]. $E_{TM}^{min} = 4.30$eV is validated by results from this work that show no photoemission from LEDs with $\lambda = 310$nm and minimal photoemission, under direct illumination only, for LEDs with $\lambda = 295$nm. We validate $E_{TM}^{max} = 4.90$eV with information from references 37 and 38 as well as results showing that UV-LEDs with $\lambda = 240$nm do not significantly enhance the emission from gold surfaces[36,39].

Explicitly considering the energy distributions of the LEDs and the TM emission sites, $I_{\lambda i}$, the photoelectric current generated by the UV-LED of central wavelength $\lambda$ is:

$$I_{\lambda i} = e \int_{E_{\lambda i}^{min}}^{E_{\lambda i}^{max}} \Phi(E)_{\lambda i} \left[ \int_{E_{TMi}^{min}}^{E_{\lambda i}^{max}} \mathbb{D}(E')_{TMi} \psi(E')_{TMi} dE' \right] dE \qquad (8)$$

where $i$ represents the $i^{th}$ reflection and, where again for generality, we assume that, due to the illumination of different areas of the TMs, the reflectivity and photoemission characteristics vary with each reflection and therefore with each $i$. The integrals are in the energy domain, with the first one over the entire range of all LED emitted energies ($E_{\lambda i}^{min}$ to $E_{\lambda i}^{max}$) and the second one from the minimum TM photoemitting energy ($E_{TMi}^{min}$) to the maximum LED-UV energy ($E_{\lambda i}^{max}$). $\Phi(E)_{\lambda i}$ is the LED photon flux at $E$, $\mathbb{D}(E')_{TMi}$ and $\psi(E')_{TMi}$ are respectively the density of photoemitting states and the quantum efficiency at $E'$ of the gold coatings. The second integration range of equation 8 is over the maximum kinetic energy of the photoelectrons; $\mathbb{K}_{\lambda i}^{max} \leq E_{\lambda i}^{max} - E_{TMi}^{min}$, as consistent with equation 1.

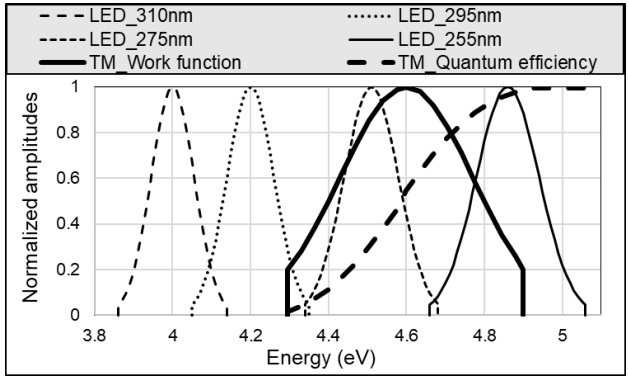

**Figure 2. Modeled power profiles of LEDs, and of work function and quantum efficiency for TMs**

Figure 2 is a plot of the power profiles of the LEDs used in this work based on the manufacturer's specifications and where we approximate a triangular profile as a gaussian distribution with cutoffs. We also show the modeled energy profile for the work function and its integral (quantum efficiency) for the gold coatings of the TM; with the parameters specified in Table 2 (with cutoffs). All amplitudes are normalized to unity. The overlap of the profiles of each of the LEDs and the quantum efficiency of the TM (integral of work function) indicates the kinetic energy range of the photoelectrons. $\mathbb{V}_{TM,\lambda}^{max}$, the maximum TM potential under illumination by an LED of $\lambda$ wavelength and surrounded by a grounded enclosure is:

$$\mathbb{V}_{TM,\lambda}^{max} \leq \mathbb{K}_{\lambda}^{max}/e \leq (E_{\lambda}^{max} - E_{TM}^{min})/e \qquad (9)$$

**Table 3. $\mathbb{V}_{TM,\lambda}^{max}$ for the various LEDs**

| LED_λ (nm) | 310 | 295 | 275 | 255 |
|---|---|---|---|---|
| $\mathbb{V}_{TM,\lambda}^{max}$ (mV) | -165 | 48 | 380 | 760 |

Table 3 gives $\mathbb{V}_{TM,\lambda}^{max}$ for all values of $\lambda$ obtained by using the models for the TM and the LEDs shown in Figure 2. We did not detect photoemission for $\lambda = 310$ nm: $I_{310} < 10$fA. As the main goal of this work is the study of the equilibrium TM potentials after exposure to UV, with no bias applied, we only measured the photoemission efficiency for the 255 nm LED. The photoemission efficiency at other wavelengths and the dependence of the photocurrent on the difference between photon energy and work function[36] will be presented in a future publication.

### IV. The Fast Photoelectron Method (FPE)

In this work we describe experiments addressing the capability of discharging for both TM charging directions. This is particularly important as $V(TM^{TE})$, the triboelectric potential of the TM upon uncaging, can be of either polarity and quite large; for example, for GP-B[12] $V(TM^{TE}) > 400$mV. Charging by cosmic radiation, $dQ(TM^{CR})/dt$, for LPF[40] and LISA[8,41] has been shown by calculation and confirmed by flight data[10,12] to be positive and, in the absence solar flare particles, $dQ(TM^{CR})/dt \cong 100\ e^+/s \approx 20$aA or $dV(TM^{CR})/dt \approx 50$mV/day. This translates to $t_{dis}^{CR}$, the time between discharges (the charge can be controlled between $\pm V_{TM}^{max}$):

$$t_{dis}^{CR} \equiv 2 \times V_{TM}^{max}/[dV(TM^{CR})/dt] \cong 3\ days \tag{10}$$

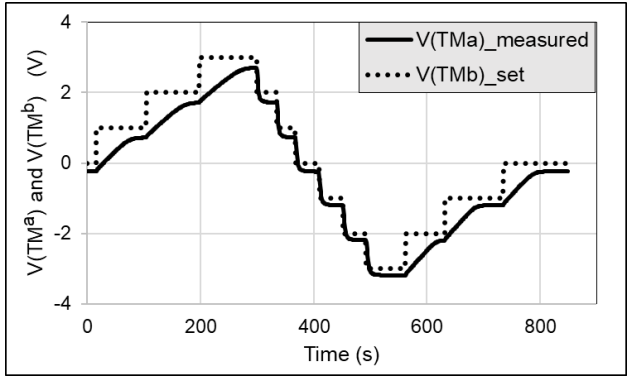

Figure 3. A typical experimental run

For comparison, $I^{PE}$, the photoelectric current generated by the UV-LEDs, at a typical operating current of 10 mA and average quantum efficiency of $10^{-6}$, is $I^{PE} > 30$pA, or $2\times10^6$ times larger than the above cosmic radiation charging rate. However, this is not an appropriate mode of continuous operation for LISA, as it will cause excessive charge induced acceleration noise due to the shot noise in the photocurrent. In practice the photocurrent would be reduced to the LISA requirement level for continuous operations, or the systems will be activated for short periods at 3-day intervals.

We consider the case of only one LED operating ($LED^A$ on) and $TM^a$ grounded ($V(TM^a) = 0V$). From equation 7 follows that in order to achieve equilibrium with $V(TM^b) = 0V$ and $I^{Aa} = 0A$ we require $\Psi^a = \Psi^b R^a$. However, control of the $\Psi$ and $R$ parameters is not practical. Consequently, equilibrium for the single source FPE system (where the photoelectrons have energies above $V_{TM}^{max}$) will occur when $f(V(TM^b))$, a function of the TM potential that provides the bias for the work function and therefore the quantum efficiency $\Psi^b$, satisfies:

$$\Psi^a = (\Psi^b + f(V(TM^b)))R^a \tag{11}$$

For the gold coatings deposited for the present work on both $TM^a$ and $TM^b$ at Stanford, as well as for the coatings that Ames Research Center applied to the UV-LED flight experiment[26], this occurs at $V(TM^b) \approx 250$mV. The FPE approach to charge management is therefore based on adjusting the photon fluxes $\Psi^{a,b}$ of the two opposing LEDs, thus balancing equation 7 experimentally, so that $I^{(A+B)a} = 0A$ and $V(TM^b) = 0V$ (with $TM^a$ grounded) resulting in:

$$\Phi^A/\Phi^B = (\Psi^b - \Psi^a R^b)/(\Psi^a - \Psi^b R^a) \tag{12}$$

Table 4 gives the settings for the experimental runs for the FPE method experiments, where all LEDs used have wavelength $\lambda = 255$ nm. The left column shows the notation used for referencing a given table row. In the last column we designate the configuration of the experiment. 'Direct' Mode of Illumination (DMI) indicates that the LED is across from the monitored plate ($LED^A$ and $TM^b$, or $LED^B$ and $TM^a$) and the photoemission takes place under direct illumination.

In the 'Reflected' mode (RMI) LEDs and plates are on the same side of the system ($LED^A$ and $TM^a$, or $LED^B$ and $TM^b$) with the electrons emitted by reflected photons. For the FPE method both LEDs are actuated, and we specify this configuration as the 'Binary' Mode Illumination (BMI). With no applied bias and our system's similar gold coatings, DMI and RMI will result in positive and negative charging respectively, as the electrons are being primarily removed from or deposited on the monitored TM.

Table 4. Settings for FPE method. $\lambda(LED) = 255nm$

| Fig. | $I(LED^A)$ (mA) | $I(LED^B)$ (mA) | $TM^a$ | $TM^b$ | TM Illum. |
|---|---|---|---|---|---|
| 3 | Off | 10 | $V_B^a$ | $V(TM^b)$ | RMI |
| 4a | Off | 10, 15, 20 | $V(TM^a)$ | $V_B^b$ | DMI |
| 4b | Off | 10, 15, 20 | $V(TM^a)$ | $V_B^b$ | RMI |
| 5a | 0→12 | 10 | $I(TM^a)$ | $V_G^b$ | BMI |
| 5b | 0→12 | 10 | $V(TM^a)$ | $V_G^b$ | BMI |
| 6 | 0→16 | 0→20 | $V_G^a$ | $V_G^b$ | BMI |
| 7a-d | 8.126 | 10 | $V_G^a$ | $V(TM^b)$ | BMI |
| 8a-d | 10 | 7.5 | $V(TM^a)$ | $V_B^b$ | BMI |

We first show the results of two runs used to validate the functionality of the system. Figure 3 and Table 4.3 show a typical test run where $I(LED^B) = 10$mA. The bias voltage on $V_B^a$ is stepped at 1 V intervals from 0 V to 3 V, to -3 V, and back to 0 V, while monitoring the $V(TM^b)$ potential until it reaches equilibrium. $V(TM^a) < V(TM^b)$ for all settings of the bias potential, as expected for an RMI test. Verification runs performed while reversing the symmetry of the system, i.e., with the LED in position $LED^A$ and the roles of $TM^a$ and $TM^b$ switched, yielded similar results.

In order to confirm the effect of the magnitude of the excitation current (and thus of the photon flux that is proportional to it) on the system performance, we performed runs in which we monitor the $TM^a$ potential while operating $LED^B$ at 10 mA, 15 mA, or 20 mA. $TM^b$ is switched from 0 V to 1 V, $V(TM^a) < V(TM^b)$ (Figure 4a, Table 4.4a), then to 0 V, $V(TM^a) > V(TM^b)$, (Figure 4b, Table 4.4b). For tests 4a and 4b the applied biases cause the illumination modes to become DMI and RMI respectively, and equation 7 to reduce to equation 13:

$$I_{4a}^{Ba} = -\Phi^B \Psi^b/(1 - R^a R^b) \text{ and } I_{4b}^{Ba} = \Phi^B \Psi^a R^b/(1 - R^a R^b) \tag{13}$$

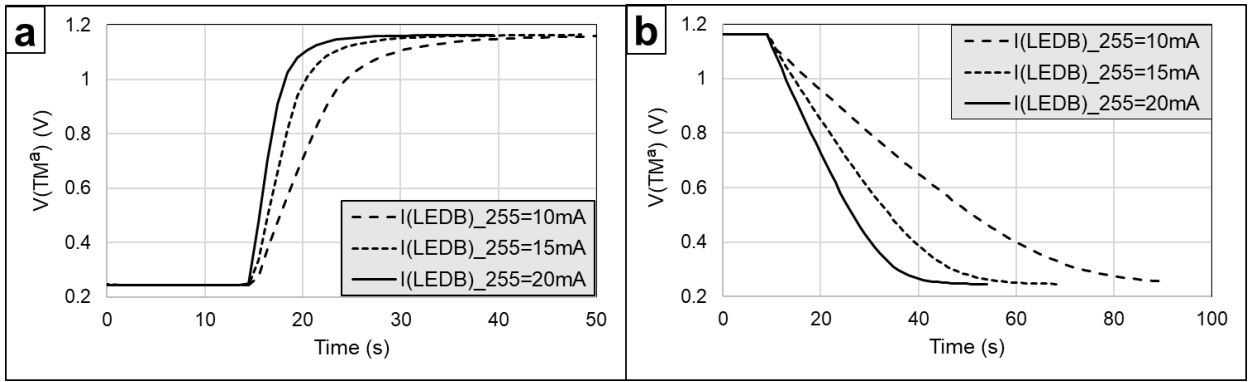

**Figure 4.** TM potential for LED_255 currents 10, 15, and 20 mA. *Left*: Bias 0 V to 1 V. *Right*: Bias 1 V to 0 V

The shorter time constant in 4a relative to 4b indicates that $\Psi^b > \Psi^a R^b$ in equation 13. As expected from the discussion in section III, the time constant to equilibrium increases with lower photon flux while the equilibrium voltage is independent of it.

**Table 5. Maxima of voltage variations and currents for the charging and discharging data from Figure 4.**

| $I(LEDB)$ (mA) | $(dV/dt)^+$ (mV/s) | $(I_{4a}^{Ba})^{max}$ (pA) | $(dV/dt)^-$ (mV/s) | $(I_{4b}^{Ba})^{max}$ (pA) |
|---|---|---|---|---|
| 10 | 100 | 16.9 | -17.0 | -2.87 |
| 15 | 162 | 27.4 | -28.9 | -4.88 |
| 20 | 230 | 38.9 | -41.7 | -7.05 |

Table 5 gives the maxima of the voltage variations, measured with $\cong 3\%$ accuracy, for the three different excitation currents $I(LEDB)$, at the start of the charging [$(dV/dt)^+$ and Figure 4a] and of the discharging [$(dV/dt)^-$ and Figure 4b] operations respectively. The corresponding $TM^a$ currents are $(I_{4a/4b}^{Ba})^{max} = C(TM^a)(dV/dt)^{+/-}$; where $C(TM^a) = 169$pF, the total $TM^a$ capacitance to the bias structure (see Table 1). Linear zero intercept fits of $(I_{4a/4b}^{Ba})^{max}$ to $I(LEDB)$ give:

$$(I_{4a}^{Ba})^{max} = (1.87 \pm 0.06) \times 10^{-9} I(LEDB) \tag{14a}$$

$$(I_{4b}^{Ba})^{max} = (-0.34 \pm 0.01) \times 10^{-9} I(LEDB) \tag{14b}$$

As $TM^a$ and $TM^b$ have identical coatings, we can set equal their photoelectric efficiencies $\Psi^a = \Psi^b = \Psi$ and reflectivities $R^a = R^b = R$, resulting in:

$$(I_{4a}^{Ba})^{max} = -\Phi^B \Psi/(1 - R^2) \text{ and } (I_{4b}^{Ba})^{max} = \Phi^B \Psi R/(1 - R^2) \tag{15}$$

The reflection coefficient is then obtained from 14a, 14b and 15:

$$R = (I_{4b}^{Ba})^{max}/(I_{4a}^{Ba})^{max} = (0.34 \pm 0.01)/(1.87 \pm 0.06) = 0.18 \pm 0.01 \quad (16)$$

This value agrees with $R = 0.17$ measured directly for our gold coating samples by a different method[30,42].

From equation 15 we calculate $\Psi$:

$$\Psi = (hc/e\lambda)[(I_{4a}^{Ba})^{max}/P_{ph}](1 - R^2) \quad (17)$$

where $\lambda = 255$nm, $P_{ph}$ is the UV photon flux that is proportional to the excitation current with $P_{ph}/I(LEDB) = 0.015$ W/A (0.3 W at 20 mA)[33], and $h, c$, and $e$ are the Planck constant, velocity of light, and electric charge. Using the values from relations 14a and 16 we obtain:

$$\Psi = (hc/e\lambda)[(I_{4a}^{Ba})^{max}/I(LEDB)](I(LEDB)/P_{ph})(1 - R^2) = (5.9 \pm 0.2) \times 10^{-7} \quad (18)$$

The value of $\Psi$ is consistent though higher than $\Psi(sample) = (3.4 \pm 0.35) \times 10^{-7}$, the value obtained for the gold samples[30,42]. The difference could be caused by the much longer pumping time for the present measurements.

For the FPE approach, both 255 nm UV-LEDs are activated. By adjusting their excitation currents, the photon fluxes of the two LEDs are varied in order to zero the potential of the TM being monitored (see equations 6 and 7). The FPE approach only works if the two LEDs generate opposite charging polarities of the TM. Note that the situation in flight will always be uncertain, and maybe different from ground-testing, as it is not possible to absolutely guarantee surface properties over the several years' timeframes involved. Flexibility is key to allow situations to be adapted as necessary, as in the LPF mission. The original LPF concept included an approach similar to the BMI tests, i.e., with photon flux variations - although also including TM charge measurement. However, as both LED sources charged the TM the same polarity, different charge management approaches were used for the LPF in flight[14]. While in this work we balance the two photoelectron fluxes by reducing the excitation current of the LED generating the higher flux, similar equilibrium can be achieved by varying the duty cycles of the LEDs.

In a first set of FPE runs the photon flux of $LED^B$ is kept constant, $I(LED^B) = 10$mA, the flux of $LED^A$ is varied, $0\text{mA} \leq I(LED^A) \leq 12$mA, and $TM^b$ is grounded. We measure the current to ground of $I(TM^a)$ (Figure 5a, Table 4.5a), or its potential (Figure 5b, Table 4.5b). The LED photon flux is linear in excitation current for $I(LED^A) \geq 4$mA, and therefore it follows from equation 7 that $I(TM^a)$ is linear in $I(LED^A)$ (Figure 5a and equation 11). However, $V(TM^a)$ is only approximately linear for a very narrow range, ± 25 mV (Figure 5b), as its potential difference to $V_G^b = 0V$ counters and saturates the current flow. This effect was not taken into consideration in section III, since we are only interested in the narrow range around the null potential. We obtain the relationships:

$$I(TM^a)(pA) = 9.05 - 1.31 \times I(LED^A)(mA) \quad (19)$$

$$V(TM^a)(mV) = 184.4 - 27.0 \times I(LED^A)(mA) \quad (20)$$

Consequently, $I(TM^a) = 0$ at $I(LED^A) = 6.93 \pm 0.02$mA, and consistently, $V(TM^a) = 0$ for $I(LED^A) = 6.83 \pm 0.10$mA (with $I(LED^B) = 10$mA). This validates the FPE method with 255 nm LEDs. The slope of the potential variation relative to the driving current change is $d(V(TM^a))/d(I(LED^A)) \cong 27$mV/mA. This means that in order to maintain a potential stability of 1 mV (or $\cong 1\%$ of $V_{TM}^{max}$), the stability of the current needs to be 40 μA, an easily achievable goal; assuming a 1 K temperature stability. From Figure 5b we observe that $V(TM^a) \cong 0.2$V when $I(LED^B) = 0$mA, consistent with the UV-LED mission flight results[26]

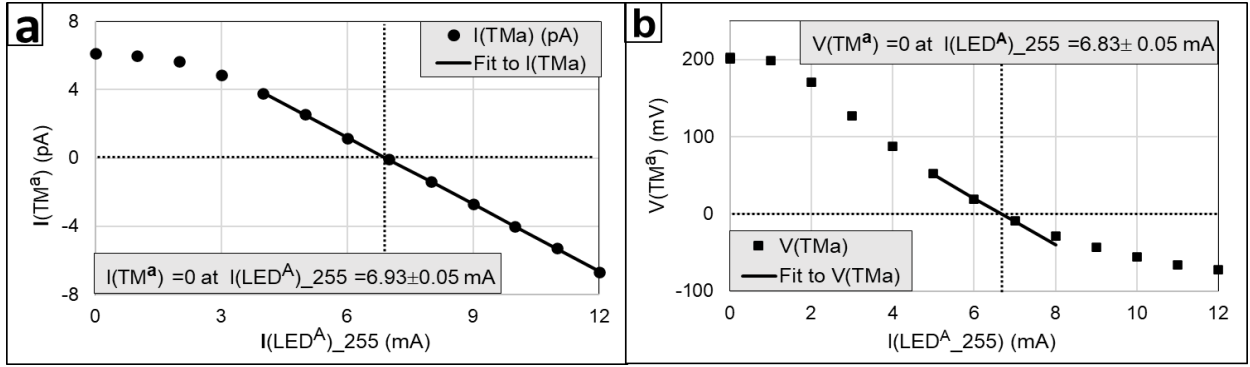

**Figure 5.** $I(TM^a)$ current to ground (*a*) and $V(TM^a)$ potential to ground (*b*) as function of $I(LED^A)$.

We next show that for FPE, in order to keep the $TM^a$ potential $V(TM^a)$ at zero (and thus the current $I(TM^a)$ at zero), the ratio of the excitation currents of the LEDs, in positions $LED^A$ and $LED^B$, needs to remain constant; in agreement with equations 6 and 7. Here we keep the measured photocurrent to ground (and thus its potential) at zero $I(TM^a) = 0$ by adjusting the ratio of the excitation currents $I(LED^A)$ and $I(LED^B)$ (while $TM^b$ is connected to the $V_G^b$ ground). The results are shown in Figure 6 and Table 4.6. One run was performed 1 hour after pump-down, while the second one was completed 21 days after vacuum recycle. In both cases and over the entire range of excitation currents, $I(LED^A)/I(LED^B)$ = constant (see constant values in Figure 6). verifying once more equations 6 and 7 and the considerations in section III. However, while the ratio of excitation currents is constant in both cases, its change of slope dependent on pumping time emphasizes the need to stabilize the system's parameters in order to achieve reproducibility of results.

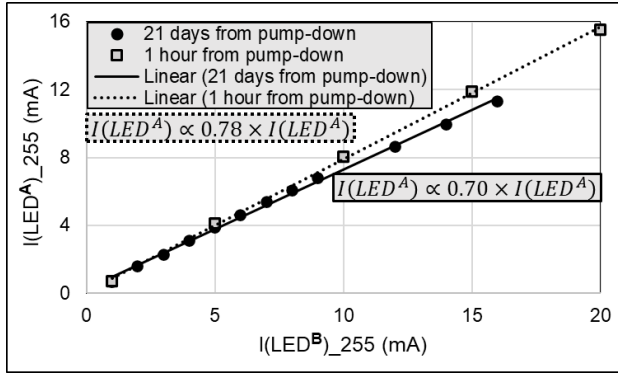

**Figure 6.** $I(LED^A)$ *vs*. $I(LED^B)$ **required for nulling the** $TM^a$ **current to ground while** $TM^b$ **is grounded. All current measurements are ± 0.05 mA.**

Figure 7 and Table 4.7(a-d) show the results and settings for a 3.5-day FPE run with 256 identical discharge cycles of 20 min each and with $V(TM^a)$ grounded: $V(TM^a) = 0V$. Sampling rate is 1 Hz. For half of each cycle both LEDs are switched on for 10 min, with the currents $I(LED^A)$ and $I(LED^B)$ set to null $V(TM^b)$, while for the following half of each cycle the LEDs are off for 10 min, while $V(TM^b)$ is allowed to drift. For this test $I(LED^A) = 8.126$ mA and $I(LED^B) = 10.0$ mA. Figure 7a shows the $V(TM^b)$ potential at the end of the drift periods with a scatter of $\leq |10mV|$ and a clear diurnal variation of $\geq \pm 5$ mV attributed to the sensitivity to temperature of the electronics. Figure 7b displays the $V(TM^b)$ potential for three 20-min cycles demonstrating the details of the FPE method. The FPE charge control system time constant is $\tau_{FPE}^{ini} = 3.0 \pm 0.5$ sec. which is dominated by instrumental limitations. The transient behavior with an initial overshoot of $\approx 2$ mV, after turn-on of the LEDs, is possibly due to thermal effects from operating the LEDs in a vacuum and is being further investigated.

Figure 7c and Figure 7d show respectively the $V(TM^b)$ potentials for the 256 discharge cycles at 3 min and 10 min after the start of the discharge; that is after $3 \times \tau_{FPE}^{equ}$ ($\tau_{FPE}^{equ}$ ≡instrumental equilibration time for the FPE method) and at the turnoff of the LEDs. Their average difference is $|V(TM^b)^{3min} - V(TM^b)^{10min}| = 0.05 \pm 0.25$mV, that is, consistent with zero and with a standard deviation within the measurement error. The drift rate of $V(TM^b)$ is:

$$dV(TM_{FPE}^{drift})/dt \cong 1.5\text{mV/day} \quad (21)$$

As an example, this rate would allow about two months for the drift of $V(TM^b)$ to reach $V_{TM}^{max} \cong 80\text{mV}$, the LISA requirement. Note that as $[dV(TM^{CR})/dt]/[dV(TM_{FPE}^{drift})/dt] \cong 30$, the drift of the FPE method is negligeable compared to the rate of charging for LISA caused by cosmic radiation. However, applicability to the LISA drag-free sensor, as well as to any other high-performance instrument, will depend on meeting their unique requirements. Figure 7 also displays the diurnal dependence of the results, that will be shown in the next section to be caused mainly by temperature variations.

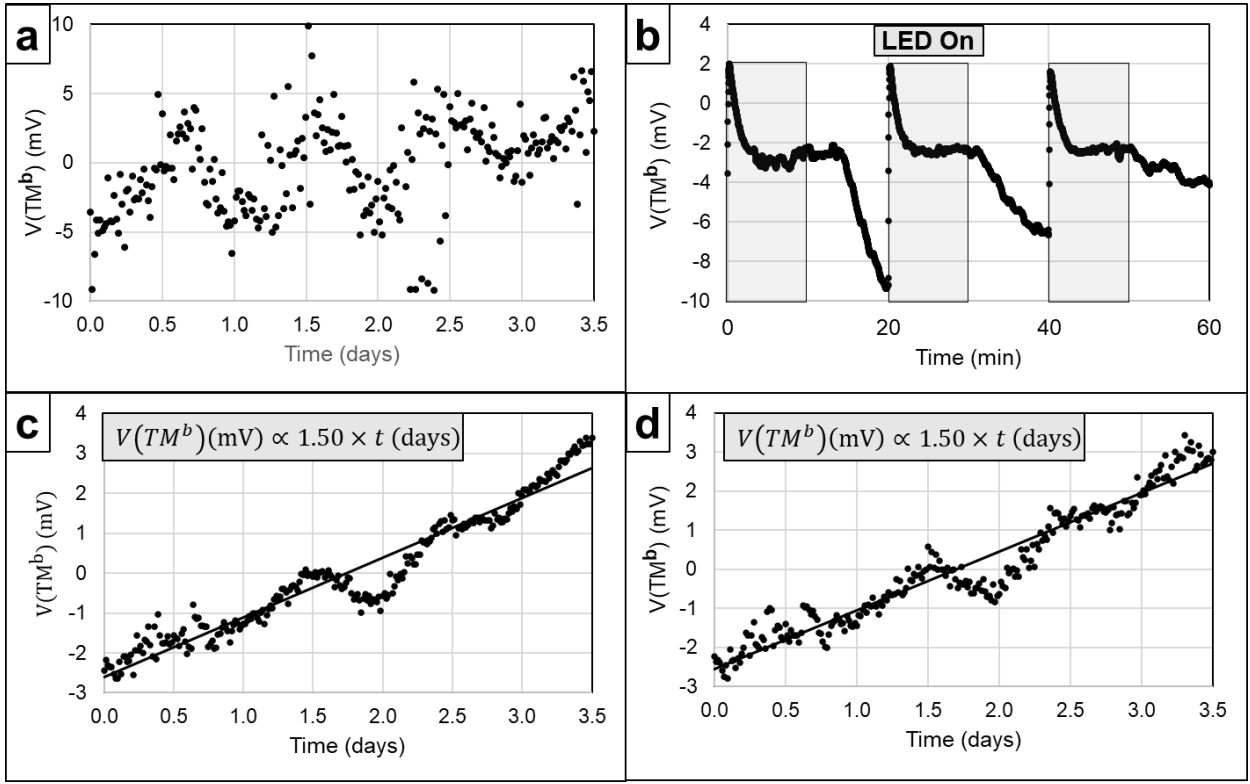

**Figure 7. Demonstration of FPE method with 10 min discharges every 20 min: 3.5 days, 256 discharges a) TM potentials at start of the discharges b) Three typical discharges showing the periods of LEDs shaded c) TM potential after 3 min of LEDs on with linear fit d) TM potential after 10 min of LEDs on with linear fit.**

Figure 8 (settings summarized in Table 4a-d) shows a verification of the FPE method for neutralization of TM potentials of ±1 V and ±2 V with emphasis on the behavior of the system for charge reduction from ±1 V to 0 V. We monitor $V(TM^a)$ while the $V(TM^b)$ potential is changed between -2 V and 2 V in steps of 1 V and the excitation currents are set to $I(LED^A) = 10.0\text{mA}$ and $I(LED^B) = 7.5\text{mA}$: Figure 8a. The $V(TM^a)$ equilibrium potentials for the five $V(TM^b)$ settings are shown in Figure 8b with $V(TM^a)/V(TM^b) \propto 0.96 \pm 0.01$. Figure 8c and Figure 8d demonstrate the detailed FPE discharge mechanism from +1 V and -1 V respectively. Initially, while the potential difference dominates the process, the discharge regime is linear with a rate of $|dV(TM^a)/dt| = 73 \pm 1 \text{ mV/s}$. The discharge current is 12±1 pA (the total capacitance to ground of $TM^a$ is 169 pF). From around 200 mV to ≅0 mV the discharge regime is exponential with a time constant of 3.3±0.2 s, consistent with the results of the tests shown in Figure 7.

The results of the FPE test illustrated in Figure 8 were validated by using the symmetry of the system and reversing the functions of the plates: the $V(TM^b)$ potential is being measured while the $V(TM^a)$ potential is being changed between -2 V and 2 V in steps of 1 V. The required currents for this arrangement are to $I(LED^A) = 10.0$ mA and $I(LED^B) = 5.9$ mA, with results for $|dV(TM^b)/dt|$ and the exponential decay time constants consistent with the reverse experiment,

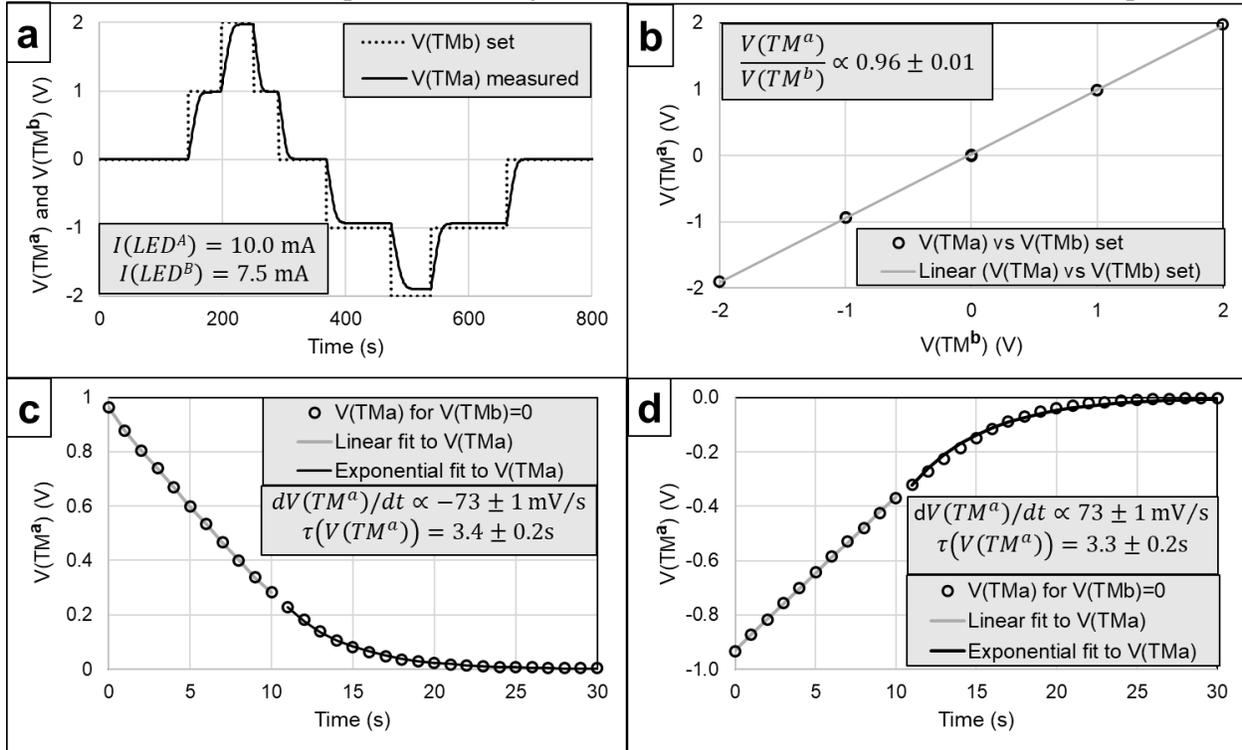

**Figure 8.** Demonstration of the FPE charge management method for voltages of $\pm 1$ V and $\pm 2$ V. a) $V(TM^a)$ as a function of time, b) $V(TM^a)$ as a function of $V(TM^b)$, c) 1 V discharge with linear and exponential fits, d) -1 V discharge with linear and exponential fits

when scaled to the $TM^b$ capacitance to ground.

The FPE experiments demonstrate that this charge management method meets LISA discharging level requirements for sensors with geometry that allows mounting of the LEDs in both direct and reflected illumination modes. The FPE configuration can be easily established and maintained, as the TM potential varies by only 1 mV for each 40 µA change in the excitation current of the 255 nm LEDs: an easily achievable current precision for tuning the system. The charge management time constant can be varied as required by modifying the currents of the two LEDs while maintaining constant their pre-established ratio. For LED currents of the order of 10 mA, the active discharge duration is less than 30 sec from $\pm 1$ V. Drift rates in the FPE mode (attributed mainly to system temperature) are $dV(TM)/dt \cong 1.5$ mV/day, well below the $dV(TM^{CR})/dt \approx 50$ mV/day charging rate due to cosmic radiation for LISA. The FPE method was developed for the Modular Gravitational Reference Sensor, (MGRS)[43], that has a simple geometry and uses no electric fields.

Adapting the FPE method to LISA will require running tests with the actual complete sensors and their electronic systems and solving compatibility issues including: a) the limitation on the photoelectron shot noise, that could be addressed by either short activation periods at high current, during no data taking maintenance periods, or by reducing the photoelectron current to meet the shot noise requirements, b) to account for the complex geometry of the LISA drag-free sensor, integrated testing is required, c) LISA (same as LPF) sensors use signal injection and actuation control voltages that are much larger than 80 mV, and moreover, in its present configuration, most photoelectrons are released in those regions of high electric fields. AC charge management, with synchronizing LED activation times with the highest local electric fields could be a solution.

All FPE tests were performed with 255 nm LEDs. Our analysis shows that similar results can be obtained for gold coatings for all wavelengths shorter than 275 nm. We note that the FPE results validate the considerations of section III, are consistent with flight experiments using similar 255 nm LEDs[26], and corroborate the approach described by F. Yang and collaborators[28]. The FPE passive charge management method is to some extent sensitive to changes in system conditions like vacuum, temperature and, possibly aging, and therefore will require periodic recalibration.

## V. The Slow Photoelectron Method (SPE)

In this set of experiments our objective was to demonstrate a discharge method with a single LED with photoelectron energy lower than $V_{TM}^{max}$. The slow photoelectrons act essentially as a low resistance short between the two TMs when one of them is connected to a variable bias voltage. For the SPE method experiments the 255 nm UV-LED is mounted permanently in position $LED^B$, while LEDs of 275 nm, 295 nm, and 310 nm are used in turn at $LED^A$. As for the FPE mode, $TM^a$ and $TM^b$ can be interchangeably operated in 'housing' or TM mode, with the 'housing' grounded for all measurements (as will be the actual housings of all instruments under consideration). We use the value of the TM housing potential and its stability, to quantify the SPE performance. Table 6 shows the summary of the settings for all SPE runs, using the same labeling conventions as those of Figure 1 and Table 4; with the addition of the LED wavelength to the definition in the table reference column

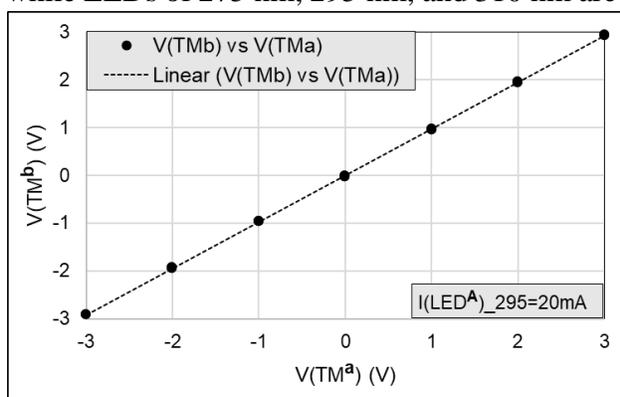

**Figure 9. TM potential for 295 nm LED SPE method.**

No photoelectric effect was observed with the 310 nm LED. The intensity of the 310 nm LED drops to about 4% of its peak power at 300 nm, which indicates that the work function of the TM gold coated surface is larger than about 4.14 eV. This result was used in section III (and also in Figure 2 and Table 2) for the model of the energy profile of the work function for the gold films. The lowest TM potential for the SPE method was obtained using the 295 nm LED in DMI mode (directly illuminating the TM): see Figure 9 and Table 6.9_295. The lack of photoemission in the RMI configuration at $\lambda = 295$nm, is caused by the minimal overlap of $E^{max}(LED\_295)$ and $E^{min}(TM)$. The fit for the data in Figure 9 is:

$$V(TM^b)(V) = 0.972 \times V(TM^a)(V) + 0.004\ (V) \tag{22}$$

Indicating that $V(TM^{b,a}) = \pm 4$mV for $V(TM^{a,b}) = 0$V.

For the series of SPE measurements shown in Figure 10a and 9b, $TM^a$ and $TM^b$ were used as either the 'housing' or the TM. Figure 10 shows the TM potential as a function of the bias potential. In Figure 10a $TM^a$ is adjusted as the bias, $V_B^a$, and the $TM^b$ potential, $V(TM^b)$, is measured, while Figure 10b has the reversed configuration. The upper left shaded inserts show a magnified version of the data points at $V_B^{a/b} = 0$V, the region of interest for the SPE method.

In the $\Delta V(TM_0)$ column of Table 6 we present the results for the experiments shown in the shaded inserts of Figure 10, indicative of the performance of the SPE method of charge management when in DMI configuration; the RMI runs are performed for system validation. As expected for the 255 nm and the 275 nm LEDs, the DMI and RMI configurations produce positive and negative offsets respectively. The three runs with the 295 nm LED in DMI configuration give similar results in agreement with $V(TM) = 0 \pm 10$mV (Figure 9 and Table 6.9_295, Figure **10**a and Table 6.10a_295, Figure 11b and Table 6.11b_295). The column 'Model $\Delta V(TM_0)$' in Table **6** gives the values of $\Delta V(TM_0)$, as estimated from the model in section III equation 1 and Table 2, that are in qualitative agreement with the measurements - as expected for our rudimentary model. For the 275 nm LED, the measurements in Figure 10a and 9b indicate that the potential is symmetric in both DMI and RMI directions. Assuming similar physical properties for the plates and reflection constants of the order of 0.5, this symmetry implies that the center of the gold surface work function is close to the maximum photon energy of the 275 nm LED. Maximum photon flux, wavelength and energy are $\Phi_{275}^{Emax} \approx 0.04 \times \Phi_{275}^{max}$, $\lambda_{275}^{Emax} = 265$nm, and $E_{275}^{Emax} = 4.68$eV, consistent with the 4.60 eV with the center of the gaussian distribution chosen for our model of the work function of the gold surfaces.

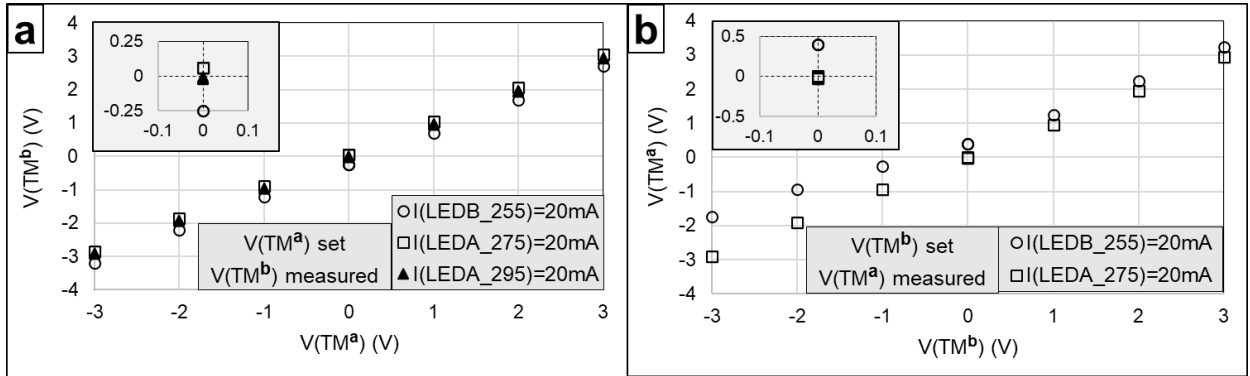

**Figure 10. TM potential as a function of the bias voltage for 255 nm, 275 nm, and 295 nm LEDs: a) $V(TM^a)$ set b) $V(TM^b)$ set**

**Table 6. Settings for SPE charge management method**

| Table Ref. | Fig. | $\lambda(LED^A)$ (nm) | $I(LED^A)$ (mA) | $\lambda(LED^B)$ (nm) | $I(LED^B)$ (mA) | $TM^a$ | $TM^b$ | Mode | $\Delta V(TM_0)$ (mV) | Model $\Delta V(TM_0)$ |
|---|---|---|---|---|---|---|---|---|---|---|
| **6.9_295** | 9 | 295 | 20 | Off | Off | $V_B^a$ | $V(TM^b)$ | DMI | -17 ± 3 | 48 |
| **6.10b_255** | 10b | Off | Off | 255 | 20 | $V_B^a$ | $V(TM^b)$ | DMI | 402 ± 2 | 760 |
| **6.10a_255** | 10a | Off | Off | 255 | 20 | $V(TM^a)$ | $V_B^b$ | RMI | -250 ± 1 | N/A |
| **6.10a_275** | 10a | 275 | 20 | Off | Off | $V_B^a$ | $V(TM^b)$ | DMI | 59 ± 1 | 380 |
| **6.10b_275** | 10b | 275 | 20 | Off | Off | $V(TM^a)$ | $V_B^b$ | RMI | -8 ± 14 | N/A |
| **6.10a_295** | 10a | 295 | 20 | Off | Off | $V_B^a$ | $V(TM^b)$ | DMI | -15 ± 6 | 48 |
| **6.11a_275** | 11a | 275 | 5 | Off | Off | $V_G^a$ | $V(TM^b)$ | DMI | 72±1 | 380 |
| **6.11b_295** | 11b | 295 | 10 | Off | Off | $V_G^a$ | $V(TM^b)$ | DMI | 0±10 | 48 |
| **6.12_295** | 12 | 295 | 10 | Off | Off | $V_G^a$ | $V(TM^b)$ | DMI | 0±10 | 48 |

To determine the stability of the SPE method, observations were performed over 5 days with the 275 nm and over 14 days with the 295 nm LEDs in position $LED^A$ and using $TM^b$ as TM and $TM^a$ as the grounded housing (Table 6.11a_275 and Table 6.11b_295). Figure 11a and Figure 11b show the potential drifts and concomitant room temperature variations for the 275 nm LED and the 295 nm LED respectively.

These results are comparable to those in Figure 7a, showing that the TM potentials vary with temperature, while their phases lag behind the room temperature due to the high thermal isolation of the EMI cage. The drift rates of the SPE potentials (days 0 to 5 for the 275 nm LED and days 6 to 14 for the 295 nm LED) are consistent with the drift rate of the FPE data for the 255 nm LED of Figure 7 and equation 13. The SPE potential values for the 275 nm LED in Figure 10a and Figure 11a are consistent (see Table 6.10a_275 and Table 6.11a_275):

$$dV_{TM}(275nm)/dt \cong 1.5mV/day \quad and \quad dV_{TM}(295nm)/dt \cong 2.0mV/day \quad (23)$$

$$V_{TM}(275nm) \cong 72 \pm 4mV \quad and \quad V_{TM}(295nm) \cong 0 \pm 10mV \quad (24)$$

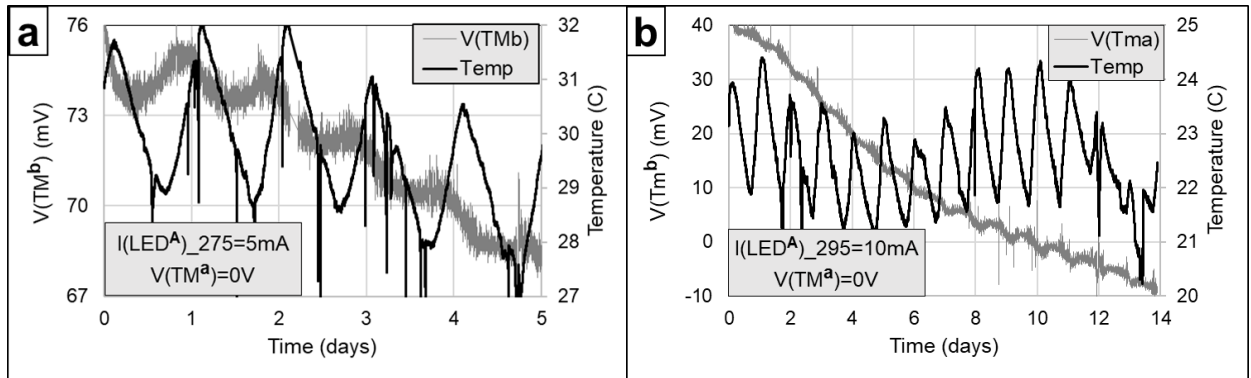

**Figure 11. TM potential drift and associated room temperature change for 275 nm and 295 nm LEDs**

Figure 12a shows the entire 14-day run for the 295 nm LED, same as Figure 11b, while also including the initial charging of $TM^b$ to 1 V and the discharging back to ground. Figure 12b displays the details of the charging and discharging occurring in the first 80 minutes of the test. The charging is performed using $LED^B$ at 255 nm and again verifies its ~200 mV equilibrium voltage when $V(TM^a) = 0V$ for the first 4 minutes in the plot.

Figure 12c presents the low-voltage discharge region of interest for actual instruments including LISA, where $V(TM^a) < 120mV$. In Figure 12d we show that this period follows, as expected, an exponential discharge law with a time constant $\tau(\lambda, I_{LED})$ given by $\tau(295, 10) = 4.3 \pm 0.2$min. This time constant was measured with the LED type MTSM295UV-F1120S at 10 mA excitation current and using a 10 cm light-pipe. $\tau(\lambda, I_{LED})$ varies inversely with $I_{LED}$ and therefore, at the standard excitation current of 20 mA $\tau(295, 20) \cong 2.2 \pm 0.1$min, requires a discharge time from 80 mV of about 5 minutes.

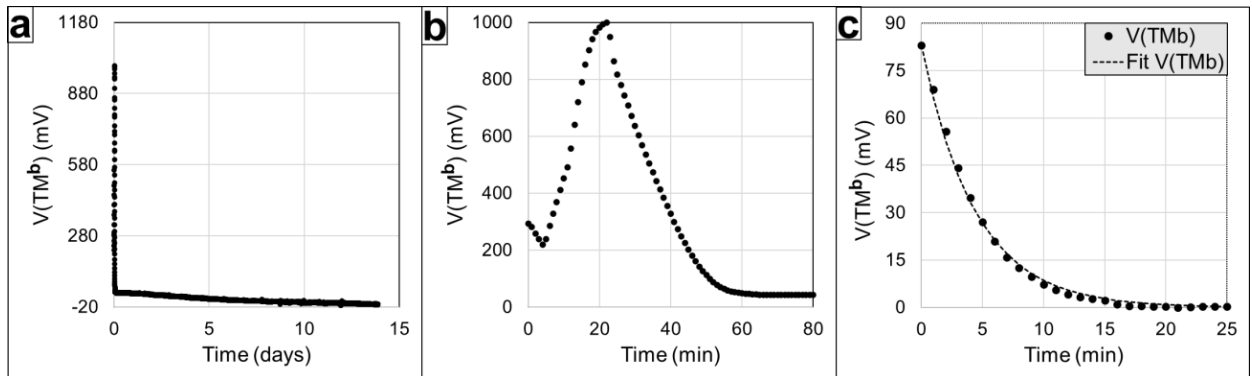

**Figure 12. a) Complete 295 nm LED 14-day run, b) expansion of section of charging to 1 V followed by discharging to ground, c) section of interest to LISA with discharging from about 80 mV; shifted to zero and with exponential fit.**

From the SPE method experimental results we determine that a single 295 nm LED in a configuration consistent with actual instrument design (directly illuminating the TM from its location on the TM housing), will maintain the TM potential at $|V(TM)| < 10$ mV, independent of the excitation current of the LED and of the system parameters, with a drift rate of the order of about 2 mV/day. The SPE results are, as required, consistent over a range of experiments, as well as with the concepts discussed in section III and the pertinent FPE results.

Our rudimentary model of the energy profile of the work function for gold films gives results consistent with the experimental ones and indicates that LEDs with wavelengths between 275 nm and 300 nm could be used for SPE method charge management.

**Conclusions**

We present two robust methods of passive charge management using photoelectrons for the floating TMs of high sensitivity instruments, that meet the goals of a passive charge management system by eliminating or greatly reducing the need for TM charge measurements and simplifying operations.

We have demonstrated a passive dual-source method that uses fast photoelectrons, FPE, generated by two adjustable intensity UV-LEDs operating at 255 nm wavelength and illuminating the TM and its housing respectively. The UV-LED intensities are adjusted by varying their excitation currents allowing easy tuning of the TM residual charge and drift rate. Following an exposure to UV of $\cong 30$ sec the system converges to zero from $\pm 1$ V with a drift rate of $dV(TM)/dt = 1.4 \pm 0.2$ mV/day. The FPE results agree with previous work[26,28]. However, periodic re-calibrations of the discharge system might be required as the FPE method is sensitive to changes in the photoemission parameters (quantum efficiency, reflectivity) and the intensity of the LEDs.

We have also demonstrated a passive bipolar charge management method using slow photoelectrons emitted under illumination by one UV LED, SPE, operating at 275-295 nm wavelength and located on the TM housing. In this regime the photoelectrons have minimal kinetic energy, and their motion is determined by the housing to TM field. Note that, like the FPE, the SPE was developed for MGRS[43] type sensors, that is, with no electric fields between TM and its housing. The same issues need to be addressed for both approaches if they are to be adapted to LISA. The slow-photoelectron system requires about 5 minutes to converge to zero from 100 mV (for $I(LED\_295) = 20$ mA) with a drift of $dV(TM)/dt = 2.0 \pm 0.5$ mV/day. Using the lowest energy photoelectrons (longest UV wavelength) possible for the SPE approach goes contrary to previous work in the field where the shortest practical UV wavelengths were used to maximize the photoelectric quantum efficiency and thus the generated photoelectric current. However, the high power of the available UV-LEDs, coupled with the low current requirements for the charge management of the instruments under consideration, makes our novel technique practical and very attractive. A single SPE type LED, mounted in the housing and illuminating the TM, will provide reliable passive charge management. Activation of the SPE system can be planned for intervals of about two days for LISA, for durations of the order of minutes. The SPE method is less sensitive to either system conditions (vacuum, temperature) or photoemission parameters (quantum efficiency, reflectivity) and variations in the intensity of the LED.

The LISA requirement for the maximum TM potential $V_{TM}^{max} \cong 80$ mV, is shown to be readily achievable, with the repeatability of passive charge control to $\leq 5$ mV.

Insight into the slow photoelectrons charge control method is provided by our simple gaussian model of the work function of gold films: 4.6 eV center, 0.2 eV standard deviation, and 4.30 eV and 4.90 eV cutoffs. We estimate that the lowest energy photoemission is obtained with the 295 nm LED at $\cong 4.35$ eV or $\cong 285$ nm, at about 4% of its maximum intensity.

**Acknowledgements**

We are grateful for the assistance provided in the laboratory by Chin Yang Lui of SN&N Electronics, Inc. and for SN&N volunteering space and staff for the work during the pandemic. We also thank Robert L. Byer of Ginzton Laboratory, Stanford University, for providing financial support.